# The nature of quantum parallel processing and its implications for coding in brain neural networks: a novel computational mechanism

Andrew S. Johnson and William Winlow

Dipartimento di Biologia, Università degli Studi di Napoli Federico II, Napoli, Italy.

**Summary**

Conventionally it is assumed that the nerve impulse is an electrical process based upon the observation that electrical stimuli produce an action potential as defined by Hodgkin Huxley (1952) (HH). Consequently, investigations into the computation of nerve impulses have almost universally been directed to electrically observed phenomenon. However, models of computation are fundamentally flawed and assume that an undiscovered timing system exists within the nervous system. In our view it is synchronisation of the action potential pulse (APPulse) that effects computation. The APPulse, a soliton pulse, is a novel purveyor of computation and is a quantum mechanical pulse: i.e. It is a non-Turing synchronised computational event. Furthermore, the APPulse computational interactions change frequencies measured in microseconds, rather than milliseconds, producing effective efficient computation. However, the HH action potential is a necessary component for entropy equilibrium, providing energy to open ion channels, but it is too slow to be functionally computational in a neural network. Here, we demonstrate that only quantum non-electrical soliton pulses converging to points of computation are the main computational structure with synaptic transmission occurring at slower millisecond speeds. Thus, the APPulse accompanying the action potential is the purveyor of computation; a novel computational mechanism, that is incompatible with Turing timed computation and artificial intelligence (AI).

**Abbreviations.** HH = Hodgkin Huxley (1952), AI = Artificial intelligence; AP = Action potential APPulse = action potential pulse

**Understanding the action potential and the action potential pulse**

The physiological action potential is the orthodox action potential described in detail by Hodgkin and Huxley (1952) (HH). The action potential was conceived as the purveyor of messages from one area of the body to another based upon electrophysiological methods of investigation and theories of computation available in the middle and late 20th century. Neural control was demonstrated to be to be frequency based on stimulation, in that higher frequencies could stimulate muscles to a greater extent, and later those sensory organs like muscle spindles and the rods and cones of the eye produced higher frequency action potentials on stimulation. Computation was still in its infancy and the prevalence of transistors in electronic circuitry led to the understandable orthodoxy that binary computation by analogy was assumed to facilitate computation. The apparent binary nature of the action potential, with its profound spike (figure 1 A), seemed to verify this philosophy. Transmission of impulses therefore became synonymous with the action potential and computation became addressed by the connections between neurons at the point they met other neurons, the synapses.

The original description of the action potential (HH) does not include a plausible or possible mechanism for propagation, latency, accuracy, or entropy. Furthermore, Cable theory (Poznanaski 2013) can only explain the opening and closing of ion channels along the membrane within limits of the available active charge (Figure 1a). What is clear from the early papers and the description of the action potential is that the rise in current (associated with voltage change) starting from resting potential is described as threshold. At this point, there is no charge placed on the leading edge of the action potential that can trigger propagation from the ion channels by electrical charge (Figure 1). Significantly, Ion channels positioned in the membrane are spaced too infrequently and too remotely for charge from one to affect another (Johnson 2015, 2018). Hodgkin and Huxley (HH) describe the current and charge flowing through the membrane from the point of threshold to the end of the refractory period, but at no point do they describe propagation or what leads to propagation or latencies.

*Synapses*. Synapses convey an action potential from one neuron to the next and studies soon showed that modulating synapses could inhibit and change the latency of connection. This is a form of simple computation, but synapses function at a slower rate than action potentials with latencies typically in the millisecond to 100ms range and can exhibit both chemical excitation and inhibition. However, they are unstable over the computational limits of the APPulse i.e., in the microseconds range (Johnson 2015, 2018, Winlow and Johnson 2021). In the retina and auditory system, synapses are separated from each other by a distance that makes communication unpredictable and thus computationally ineffective (Johnson and Winlow 2019). Conventional models using spikes and back propagation to model computation, such as Spinnaker and the Blue-Brain takes the AP peak as the quantum point of computation and circumvent timing error by using a series of inventive algorithms (Furber and Bogdan 2020, Bluebrain), rather than considering the actual mechanisms involved

**Computation between neurons** has previously been centred on the behaviour and connectivity of synapses and not the intricate nature of the pulse or the computation required in a brain neural network. Frequency computation by quantum pulses is little understood and analogy to conventional binary computation has become the orthodoxy. The action potential has therefore become the model for computation in the brain which has been assumed to act via synapse modulation both binary and analogue. This model is only sustainable for simple activity such as basic muscle control or to explain increased frequencies of neurons to sensory activity. The poor outcomes of the HH action potential and synapse computation are discussed elsewhere (Drukarch and Wilhelmus 2025, Galinsky and Frank 2025, Walter et al 2016, Winlow et al 2020, 2025, Johnson and Winlow 2017, 2018, 2019). Furthermore Tasaki (1980, 1988) had noted that a mechanical pulse accompanied the action potential, later reviewed and improved upon by Heimberg (2005) and evidenced by El Hady and Machta (2015),but many questions yet remain   However, in 2018 (Johnson and Winlow 2018), we suggested that the mechanical pulse could be maintained from the perturbations of the ion channels and explained how this provided a more plausible mechanism for propagation than cable theory (Figure 1A). The no-charge state is evident from the initiation of the action potential thus showing that the threshold holds no charge, (Figure 1B). In our view propagation of the action potential occurs by soliton mechanical forces as acting on the ion-gates (Johnson and Winlow 2018, 2020, Winlow et al 2020, Winlow and Johnson 2021) We named this combined structure the action potential pulse (APPulse) that is a logical combination of both the HH action potential and the soliton pulse. The opening of the ion gates producing the HH action potential with the ionic currents provides the entropy required for maintaining the on-going soliton thus adding the functionality of feed-forward computation. The threshold of the APPulse is the moment the ion-gates open leads to immediate refraction of the membrane until the ion concentrations re-equilibrate (Johnson and Winlow 2018). The soliton pulse thus retains its integrity by addition of energy from the opening of the ion gates. Solitons annul on collision as does the HH action potential as both, in this model, are refractory immediately after the ion channels open but will leave the soliton to progress. The threshold of the soliton is derived from the energy generated by the opening of ion channels in the nerve terminal membrane, and this becomes the physical quantum responsible for computation (Figure 1 C D and E) which must operate in the microsecond range or below as previously demonstrated in the retina (Johnson and Winlow, 2019).

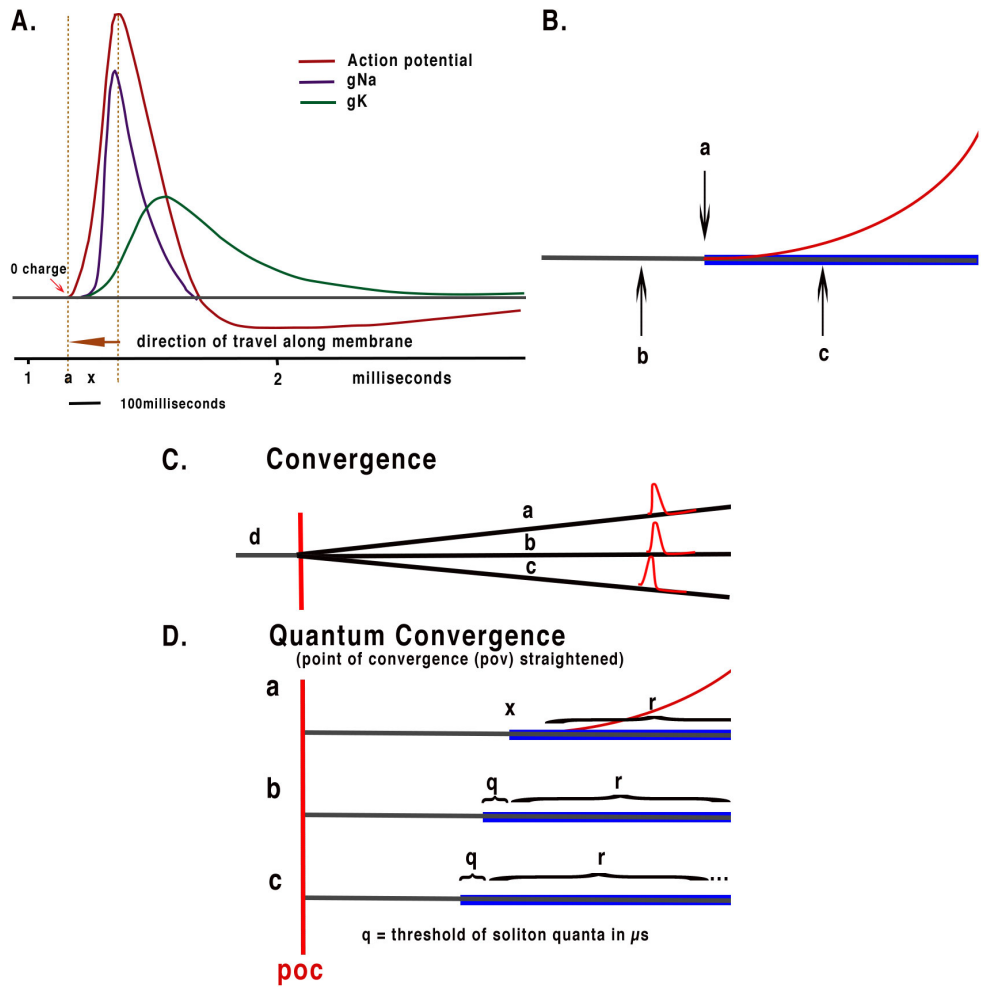

Figure 1. A. Simplified Hodgkin Huxley action potential showing beginning of charge from ion channel release and the part of membrane affected by propagation at 0 relative charge. B illustration of detail of threshold of HH action potential with membrane shown in black, mechanical 'soliton' pulse in blue and charge in red. C. Illustration of convergence of neurons without synapses. a, b and c axons converge with output to d, this type of convergence is ubiquitous in brain neurons with convergence often taking place on the soma of neurons. The point of convergence is marked in red: note that this process is forward feeding with interference between APPulse taking place at the point of computation. D is an illustration of quantum processing. a, b and c represent the same converging axons as in C with the point of computation extended for illustrative purposes only with the distance of the point of computation (POC) infinitesimally small. The beginning of the soliton marks the propagation of the action potential and is the only functioning element and can be assumed a quantum: marked as point x on neuron a. The effective quantum is illustrated in D b. where the quantum represented by q is equivalent to the abrupt portion of the mechanical 'soliton' pulse which opens the ion channels to begin the upward charge of the HH action potential with the remainder of the soliton mirroring the refractive state r of the membrane. D. c. illustrates a third APPulse a single quantum in advance of D b and two quanta beyond D a. D c as it reaches the point of computation first will annul both D a and D b effectively producing a single output.

***Computation by the APPulse.*** Both the soliton pulse and the action potential have inherent similarities and they both have a threshold for computation. Both have refractory states with the result that two colliding APPulses will annihilate as in the HH experiments.   We have concluded that measurement of the physiological HH action potential is inherently unstable with precision only in the $10^{-4}$ s range and is unreliable (Johnson and Winlow 2018, 2024), whether taken from its threshold (or the spike peak). This leads to inaccuracy of latencies and speed whenever measured (Figure 1 B C and D). Timing of computation by the HH action potential is therefore restricted to tenths of milliseconds or greater which is unfeasible considering the rapidity of perceived responses. In 2019 we demonstrated a functioning circuit diagram of the retina using frequency computation of successive APPulses where timing of converging action potentials must be in the microsecond range or less, using a mathematical evaluation of the rate of mean sampling required to explain the observed input output frequencies of converging APs. Synapses are therefore relegated from this computational method to secondary computation where synapses have both excitatory and inhibitory modification of the computation on a slower timescale. As we explained in the retina (Johnson and Winlow 2019), they downgrade activity from the cones in low light in favour of the rods. Whereas the action potential acts over milliseconds and its intrinsic speed is subject to variation the APPulse threshold is fixed in the microsecond range and is unaffected by outside influences such as ionic changes. It is only affected by changes to the structure of the membrane remaining stable over the timescale of the APPulse (Winlow and Johnson 2020). This is because the APPulse threshold is defined by the abrupt beginning of the soliton pulse flowing along a fixed membrane whose composition does not change (Heimburg 2005), whereas the action potential alone is susceptible to changes in membrane diameter, thickness, composition, and the surrounding medium. It is this precision of the APPulse that forms the basis of fixed latencies of transmission when measurements are taken from the beginning of the 'soliton threshold'. The threshold and latency are therefore defined by the soliton and not the physiological action potential. This difference is of importance when considering computation. Note also a parallel APPulse theory can explain the latencies in myelinated nerves (Johnson and Winlow 2018).

### Quantum APPulse and computation in the brain.

As can be seen from the above, the speed and therefore latency of AP is fixed to the properties of the membrane as per experimental studies while collisions between converging AP cancel according to the properties of both HH and the quantum pulse structure. Quanta have a separation of at least $10^{-6}$ second (Johnson and Winlow 2019), in other words separation must cover less than $10^{-6}$ of a metre when the APPulse travels at 1m/s. When collisions occur, computation takes place using both the quantum and the subsequent refractory period which blocks subsequent APPulses, i.e., it is frequency modulated ternary computation using threshold, membrane refraction, and time. ( Figure 2).

### 1. Cancellation of AP quanta iii and iv.

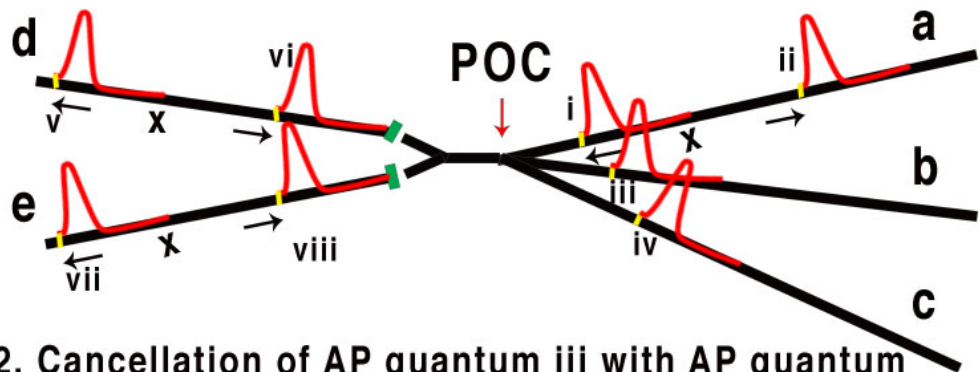

### 2. Cancellation of AP quantum iii with AP quantum iv delayed by $q+^n$.

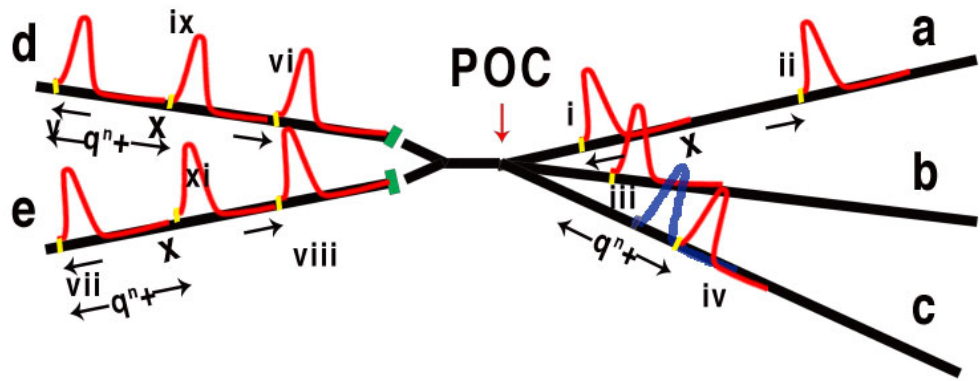

**Figure 2.** Representation of three axons (right to left) converging to a point of computation POC, synapses in green. In 1, 1. x is the distance/time between concurrent APs along axon a with a quantum shown in yellow. x is the relative latency between AP i and ii. AP iii and iv are cancelled due to overlapping the refractory membrane caused by leading AP i resulting in v, vi, vii and viii. 2. AP iv is retarded by $q+^n$ from leading quantum i where n is the timing of quantum q (see figure 1), leading to iv passing POC without cancellation resulting in additional AP ix and xi , which is a demonstration of computation and change of phase/frequency of a, b, and c.

The quantum, or threshold as it was known for the Hodgkin Huxley action potential, must be less than the frequency of an AP. Any two lesser frequencies combining will have the same quantum value; this is analogous to quantum entanglement, as each quantum has the same value to computation. In other words, two quanta one of 1μs and one 0.5μs are equal. This adaptation and combination of both the HH and Heimburg\Tasaki theories is essential to explain the higher and previously unconsidered functioning of neurons such as computation especially within a parallel brain neural network. Consequently, computation within the brain neural network is facilitated by the leading edge of the mechanical 'soliton' pulse and not the Hodgkin Huxley action potential which only provides the necessary energy for onward propagation. Computation within a nervous neural network is intrinsically linked to frequency, a fundamental aspect of this is the latency between points of contact. (Johnson and Winlow (2018) described how quanta redact other quanta in computational terms. In a network filled with colliding quanta, frequencies will change according to the changing of input frequencies as the quanta either combine or redact as demonstrated in the retina (Johnson and Winlow 2019). Note that the retina is a multi-layered forward feed network allowing cross-calculations facilitated by horizontal cells and amacrine cells where any form of synapse-oriented computation is negated by critical timing, before successive outputs are coded in parallel by the optic nerve to the Lateral geniculate nucleus. Conventional computing using timed Turing processing is incompatible with the activity and computation within the brain because both AI and quantum computing have both different logic and mechanical components compared with computation in the brain. Frequency based computation therefore appears to be the fundamental computational method used in nature. Thus, there is no mechanism for either central timing to effect synchronisation nor adequate time for synapses to affect other synapses before perception in real time. Quantum computation in neurons takes place at the speed and precision of the APPulse quantum with accuracy in the microsecond range. By extension all other computation in the brain must act similarly as the histology of neuronal convergences and synapse location are ubiquitous. Almost all neuron - synapse to neuron connections are direct and change connectivity by changes in latency. The nature of the quantum is one of a pressure-wave able to open ion channels in order facilitate propagation of the action potential and thus provide ongoing entropy for the soliton (Johnson and Winlow 2017, 2018, 2019).

The APPulse is essential to neural computation and it explains how propagation occurs at the molecular level with the speed of ion channel opening and closing and the concomitant 'soliton pulse' facilitated in microseconds. The APPulse redacts with precision at each-and-every APPulse interference, producing theoretically an unlimited number of concise error free parallel inputs and outputs into the network – this is impossible with conventional computing. Synchronisation and memory are formed simultaneously within non-defined loops within the network cortex areas by loops formed from the network (Winlow et al 2020, 2021, 2025). Many of these synchronised loops have recently been evaluated such as cortico-thalamic loops (Winlow and Johnson 2020) and we have discussed memory elsewhere (Winlow et al 2020, 2025, Johnson and Winlow 2017, 2018, 2019, 2021). Loop connections create an active circulation of AP along two or more neurons where synapses reconnect. That such mechanisms can be functionally described by the components of the APPulse is itself evidence of its accuracy, and we have extended this theory to include the perception of objects and the elucidation of synchronicity in the whole brain (Winlow et al 2023,).

Synapses are an essential part of this process. They separate cellularly diverse structures where each neuron, its shape morphology and physical location have a functional basis. This redistribution of circulating synchronised memory is essential in learning and is predetermined by the plasticity inherent in the synapses. We suggest (Johnson and Winlow 2024) that synapses act as changers of latency in an inherently plastic environment. Thus, a brain neural network is by no means fixed in functional form but adjusts according to activity.

Computation is therefore frequency-based and dependent on the length of the refractory period being many times greater than that of the quantum. Computation through a parallel network is not therefore time dependent but is dependent upon the frequency changes of the signal quantum, error free at each stage. Thus, there is no common element that exists between the evidence of nerve transmission and current conventional computational models, or indeed in computational science. The importance of error management cannot be understated as it permits lossless signalling of quantal information along parallel neurons.

**Conclusions.**

Until now, frequency modulated computation has not been considered by computer scientists and has followed conventional almost exclusively binary and 'clocked' computation. Their assumption that commercial success defines the workings of the brain is therefore unsupportable. We propose that the APPulse and quantum frequency computation is the correct neural mechanism for computational physiology of nervous communication and provides far greater efficiency and effectiveness for computation as it eradicates error that exists within in-silico and other Turing based systems.

- It is clear from the evidence that HH cannot account for the precision needed to facilitate parallel computation in the brain. The flawed assumption that the brain is a Turing machine is based upon accurate timing and synchronising the whole neural network with a centralised clock has no basis. Using the APPulse a central clock is unnecessary because computational events synchronise during frequency computation across parallel threads.
- The quantum element of the APPulse takes the form of the mechanical opening of the ion-gates at the point of threshold by the pulse and is essential for synchronised parallel computation.
- The HH action potential is therefore a mechanism for entropy exchange enabling the pulse to continuously propagate along the membrane with the ion channels opening on mechanical stimulus from the leading edge of the soliton.
- The novel computational systems of frequency-based quantum processing we have described are ubiquitous in biological neural networks. Network Parallel computation is similar to what occurs in conventional parallel circuits except that the quantum is defined in ternary terms, where the quantum and the refractory period are +1 and -1 and timing denotes whether inputs are 0.
- Conventional contemporary AI and quantum computation are not new technology but merely an attenuation of Turing technology by using algorithms that better select similarities by probability matching. They bear no relation to the computational method of the brain,

References

- Hodgkin AL and Huxley AF. "A quantitative description of membrane current and its application to conduction and excitation in nerve". Journal of Physiology 117.4 (1952): 500-544.
- Johnson A. S. "The coupled action potential pulse (APPulse)-Neural network efficiency from a synchronised oscillating lipid pulse Hodgkin Huxley action potential". EC Neurology 2 (2015): 94-101.
- Johnson A. S and Winlow W. "The soliton and the action potential: primary elements underlying sentience". Frontiers in Physiology 9 (2018): 779.
- Johnson A. S and Winlow W. "Computing action potentials by phase interference in realistic neural networks". EC Neurology 5.3 (2017): 123-134.
- Johnson AS and Winlow W. "Are neural transactions in the retina performed by phase ternary computation?" Annals of Behavioural Neuroscience 2.1 (2019): 223-236.
- Winlow W and Johnson A. S. "Nerve impulses have three interdependent functions: Communication, modulation, and computation". Bioelectricity 3.3 (2021): 161-170.
- Johnson A. S and Winlow W. "Does the brain function as a quantum phase computer using phase ternary computation?" Frontiers in Physiology 12 (2021): 572041.
- Winlow W and Johnson AS. "The action potential peak is not suitable for computational modelling and coding in the brain". EC Neurology 12.4 (2020): 46-48.

- Iwasa, I. Tasaki, and R. C. Gibbons, "Swelling of nerve fibers associated with action potentials," Science 210(4467), 338–339 (1980).
- Tasaki, "A macromolecular approach to excitation phenomena: chanical and thermal changes in nerve during excitation," Physiol. Chem. Phys. Med.NMR 20(4), 251–268 (1988).
- Tasaki, K. Kusano, and P. M. Byrne, "Rapid mechanical and thermal changes in the garfish olfactory nerve associated with a propagated impulse," Biophys. J. 55(6), 1033–1040 (1989).
- Heimburg T and Jackson AD. "On soliton propagation in biomembranes and nerves". Proceedings of the National Academy of Sciences of the United States of America 102.28 (2005): 9790-9795.
- Johnson A. S, Winlow W. Neurocomputational mechanisms underlying perception and sentience in the neocortex. Front Comput Neurosci. 2024 Mar 5;18:1335739. doi: 10.3389/fncom.2024.1335739. PMID: 38504872; PMCID: PMC10948548.
- Winlow W.,et al. "Classical and non-classical neural communications". OBM Neurobiology 7.3 (2023): 1-11.
- Winlow W. "The plastic nature of action potentials". In: The cellular basis of neuronal plasticity - physiology, morphology and biochemistry of molluscan neurons, Ed A.G.M. Bulloch. Manchester University Press, UK (1989): 3-27.
- El Hady A and Machta B. "Mechanical surface waves accompany action potential propagation". Nature Communications 6 (2015): 6697.
- William Winlow., et al. "Passing the Wave - How Does the Action Potential Pulse Pass Between Axons and Dendrites?". EC Neurology 17.1 (2025): 01-05.
- Drukarch B, Wilhelmus MM. Understanding the Scope of the Contemporary Controversy about the Physical Nature and Modeling of the Action Potential: Insights from History and Philosophy of (Neuro)Science. OBM Neurobiology 2025; 9(1): 269; doi:10.21926/obm.neurobiol.2501269.
- Galinsky VL and Frank LR (2025) The wave nature of the action potential. Front. Cell. Neurosci. 19:1467466. doi: 10.3389/fncel.2025.1467466
- Walter, F., Röhrbein, F., & Knoll, A. (2016). Computation by Time. Neural Processing Letters, 44(1), 103-124. https://doi.org/10.1007/s11063-015-9478-6
- Steve Furber (ed.), Petrut Bogdan (ed.) (2020), "SpiNNaker: A Spiking Neural Network Architecture", Boston-Delft: now publishers, http://dx.doi.org/10.1561/9781680836523
- https://bluebrain.epfl.ch/bbp/research/domains/bluebrain/
- Brzychczy S, Poznanski R R. Mathematical Neuroscience. Academic Press, 2013 ; ISBN, 0124104827